\documentclass[final,5p,times,twocolumn]{elsarticle}

\usepackage{graphicx}
\usepackage{amsmath}
\usepackage{amssymb}

\journal{Physics Letters B}

\begin{document}

\begin{frontmatter}

\title{First Measurements of Timelike Form Factors of the Hyperons, $\Lambda^0$, $\Sigma^0$, $\Sigma^+$, $\Xi^0$, $\Xi^-$, and $\Omega^-$, and Evidence of Diquark Correlations}

\author[nu]{S.~Dobbs}
\author[nu]{A.~Tomaradze}
\author[nu]{T.~Xiao}
\author[nu]{Kamal~K.~Seth\corref{cor1}}
\ead{kseth@northwestern.edu}

\address[nu]{Northwestern University, Evanston, Illinois 60208, USA}

\author[wsu]{G.~Bonvicini}

\address[wsu]{Wayne State University, Detroit, Michigan 48202, USA}

\begin{abstract}
Using 805~pb$^{-1}$ of $e^+e^-$ annihilation data taken with the CLEO-c detector at $\psi(3770)$, $\sqrt{s}=3770$~MeV, we report the first measurements of the electromagnetic form factors of the $\Lambda^0$, $\Sigma^0$, $\Sigma^+$, $\Xi^0$, $\Xi^-$, and $\Omega^-$ hyperons for the large timelike momentum transfer of $|Q^2|=14.2$~GeV$^2$.  The form factors for the different hyperons are found to vary by nearly a factor two.  It is found that $|G_M(\Lambda^0)|=1.66(24)\times|G_M(\Sigma^0)|$.  The $\Lambda^0$ and $\Sigma^0$ hyperons have the same $uds$ quark content, but differ in their isospin, and therefore the spin of the $ud$ quark pair. It is suggested that the spatial correlation implied by the singlet spin--isospin configuration in the $\Lambda^0$ is an example of strong diquark correlations in the $\Lambda^0$, as anticipated by Jaffe and Wilczek.  Improved measurements of the branching fractions of $\psi(2S)\to p\bar{p}$ and hyperon--antihyperon pairs are also reported.
\end{abstract}

\end{frontmatter}

Electromagnetic form factors of hadrons at large momentum transfer provide valuable insight into their quark-gluon structure.  However, except for the proton and the neutron, form factors of none of the other baryons have been measured at large enough momentum transfers to provide a sensitive look into their inner structure. 

In 1961 Cabibbo and Gatto~\cite{cabibbo} first proposed that the electromagnetic form factors of hadrons can be studied by $e^+e^-$ annihilation for timelike momentum transfers, $Q^2<0$, by measuring hadron pair-production cross sections.  They advocated the measurement of the form factors of nucleons and ``strange'' baryons, $B=p$, $\Lambda$, $\Sigma$, and $\Xi$, even before their quark structure was realized, by measuring $\sigma(e^+e^-\to B\overline{B})$.  In the present context of QCD and the quark-gluon structure of hadrons, it is particularly interesting to measure form factors of hyperons which may be expected to reveal the effects of $SU(3)$ breaking, as successively one, two, and three of the up/down quarks in the nucleon are replaced by strange quarks in ($\Lambda,\Sigma$), $\Xi$, and $\Omega$, respectively.  
The interest is further enhanced at large momentum transfer, such as $|Q^2|=14.2$~GeV$^2$ at which we make our measurements. This momentum transfer corresponds to a spatial resolution of $\sim0.05$~fm and provides deep insight into possible short-range correlations between the quarks.  Among these the most important are diquark correlations, which have been extensively discussed in the past~\cite{review}, and whose importance in low-energy QCD dyamics has been more recently emphasized by Jaffe~\cite{jaffe} and Wilczek~\cite{wilczek}. The differences in quark configurations between different hyperons make them an ideal laboratory to study such correlations, a dramatic example of which is provided by the effect of isospin difference between the $\Lambda^0$ and $\Sigma^0$ hyperons as revealed in the measurements we report.

Theoretical studies of hyperon form factors are very scarce.  In 1977, K\"orner and Kuroda~\cite{kornerkuroda} made predictions of $e^+e^-\to\gamma^*\to B\overline{B}$ cross sections for nucleons and hyperons for timelike momentum transfers ranging from threshold to $|Q^2|=16$~GeV$^2$ in the framework of the Generalized Vector Dominance Model (GVDM). These predictions were not constrained by any experimental measurements and turn out to be factors 10 to 80 larger than what we measure in this paper.  Recently, Dalkarov et al.~\cite{dalkarov} have made predictions of form factors of the $\Lambda^0$ and $\Sigma^0$ for momentum transfers from threshold to $\sqrt{s}=|Q|\approx2.4$~GeV, or $|Q^2|\approx5.8$~GeV$^2$, using a phenomenological model for the baryon-antibaryon interaction.

Prior to the measurements reported in this letter, only two experimental measurements of hyperon pair production cross sections and form factors existed in the literature.  In 1990, DM2~\cite{dm2} reported upper limits for the cross sections, of $\sigma(e^+e^- \to \Lambda^0\overline{\Lambda}^0,~\Sigma^0\overline{\Sigma}^0,~\text{and}~\Lambda^0\overline{\Sigma}^0)$ at $\sqrt{s}=2.4$~GeV, or $|Q^2|=5.8$~GeV$^2$. The only other measurement was made in 2007 by BaBar~\cite{babar} using the method of initial state radiation (ISR) to produce  $\Lambda^0\overline{\Lambda}^0$, $\Sigma^0\overline{\Sigma}^0$, and $\Lambda^0\overline{\Sigma}^0$ pairs from threshold to $\sqrt{s}=3$~GeV, or $|Q^2|=9$~GeV$^2$. Good statistical precision was obtained near threshold, but because of the very rapid $(\sqrt{s})^{10}$ fall-off of the cross sections, by $|Q^2|\approx9$~GeV$^2$, only upper limits could be set.

In this Letter, we report measurements of the form factors of charged and neutral hyperons, $B\equiv\Lambda^0,\Sigma^0,\Sigma^+,\Xi^-,\Xi^0,~\text{and}~\Omega^-$ for the timelike momentum transfer of $|Q^2|=14.2$~GeV$^2$~\cite{fn1}. These measurements constitute the world's first measurements of hyperon form factors with good precision and for a large momentum transfer.

We use data taken with the CLEO-c detector, which has been described elsewhere~\cite{cleodetector}, at $\psi(3770)$, $\sqrt{s}=3.77$~GeV, with the integrated luminosity $\mathcal{L}=805$~pb$^{-1}$. 
In order to use data taken at $\psi(3770)$ to determine hyperon form factors it is necessary to determine the strong interaction yield of the hyperon pairs at the resonance. 
We do so by using the pQCD prediction that the hadronic and leptonic decays of $\psi(nS)$ states scale similarly with the principle quantum number $n$.
This relation was successfully used by us recently to measure form factors of pions and kaons at the $\psi(3770)$ and $\psi(4160)$~\cite{cleo-ff}.  In the present case, it leads to 
\begin{multline}
\frac{\mathcal{B}(\psi(3770)\to\text{gluons}\to\text{hyperons})}{\mathcal{B}(J/\psi,\psi(2S)\to\text{gluons}\to\text{hyperons})} \\  = \frac{\mathcal{B}(\psi(3770)\to\gamma^*\to\text{electrons})}{\mathcal{B}(J/\psi,\psi(2S)\to\gamma^*\to\text{electrons})}
\end{multline}
Using the measured branching fractions for the $J/\psi$, $\psi(2S)$~\cite{pdg}, and the present work, we find that $\mathcal{B}(\psi(3770)\to\text{hyperons})<4\times10^{-7}$ for all hyperons, and they lead to the expected number of events, $1.3~p$, $0.9~\Lambda^0$, $0.2~\Sigma^+,\Sigma^0,~\Xi^-$, $0.05~\Xi^0$, and $0.03~\Omega^-$ for resonance decays of the $\psi(3770)$ in the present measurements.  In other words, the contributions of strong decays are negligibly small in all decays, and the observed events at $\sqrt{s}=3770$~MeV arise from the decays $e^+e^- \to \gamma^* \to B\overline{B}$.

We also use CLEO-c data taken at the $\psi(2S)$, $\sqrt{s}=3.686$~GeV, with luminosity $\mathcal{L}=48$~pb$^{-1}$, which corresponds to $N(\psi(2S))=24.5\times10^6$, to measure the branching fractions for the decays $\psi(2S)\to B\overline{B}$.  The large yield from resonance production of $B\overline{B}$ pairs from the $\psi(2S)$ enables us to test the quality of our event selection criteria, and to determine contributions to systematic uncertainties.

For decays at both the $\psi(2S)$ and $\psi(3770)$ we reconstruct the hyperons in their following major decay modes (with branching fractions \cite{pdg} listed in parentheses): $\Lambda^0\to p\pi^-$ (63.9\%), $\Sigma^+\to p\pi^0$ (51.6\%), $\Sigma^0 \to \Lambda^0 \gamma$ (100\%), $\Xi^-\to\Lambda^0\pi^-$ (99.9\%), $\Xi^0\to\Lambda^0\pi^0$ (99.5\%), $\Omega^- \to \Lambda^0 K^-$ (67.8\%).  We find that reconstructing back-to-back hyperons and anti-hyperons whose decay vertices are separated from the interaction point results in essentially background free spectra, as described in detail below.

Charged particles are required to have $|\cos\theta|<0.93$, where $\theta$ is the polar angle with respect to the $e^+$ beam.
We identify charged hadrons using the energy loss in the drift chamber ($dE/dx$), and the log-likelihood, $L^\mathrm{RICH}$, information from the RICH detector. 
We use the combined likelihood variable, for particle hypotheses $i,j\equiv \pi,K,p$,
\begin{small}
$$\Delta \mathcal{L}_{i,j} = [-2\ln L^\mathrm{RICH} + (\chi^{dE/dx})^2]_i - [-2\ln L^\mathrm{RICH} + (\chi^{dE/dx})^2]_j,$$
\end{small}
We identify protons by requiring that the measured properties of the charged particle be more like a proton than either a charged pion or kaon, i.e., $\Delta \mathcal{L}_{p,\pi}<0$ and $\Delta \mathcal{L}_{p,K}<0$.  Kaons from the decay $\Omega^-\to \Lambda^0 K^-$ suffer from a large combinatorial background, and we require $\Delta \mathcal{L}_{K,\pi}<-9$ and $\Delta \mathcal{L}_{K,p}<-9$. 
For the $p\bar{p}$ final state, proton event selection includes muon rejection and smaller acceptance, $|\cos\theta|<0.8$, as described in Ref.~\cite{cleo-ff}. 
To eliminate potential backgrounds from electrons, we use the variable $E_{CC}/p$, where $p$ is the track momentum measured in the drift chamber, and $E_{CC}$ is the shower energy in the calorimeter associated with the track.  Electrons have $E_{CC}/p\approx1$, and we require protons to have $E_{CC}/p<0.85$. 

Any number of photons are allowed in an event.
Photon candidates are calorimeter showers in the ``good barrel'' ($\cos\theta=0-0.81$) or ``good endcap'' ($\cos\theta=0.85-0.93$) regions that do not contain one of the few noisy calorimeter cells, are inconsistent with the projection of a charged particle track, and have a transverse energy deposition consistent with that of an electromagnetic shower.  We reconstruct $\pi^0\to\gamma\gamma$ decays by requiring that photon candidate pairs have mass within $3\sigma$ of the known $M(\pi^0)$, and then kinematically fitting them to $M(\pi^0)$.  
The $\pi^0$ candidates are initially assumed to originate from the interaction point, however the $\pi^0$ candidates used to reconstruct $\Sigma^+$ and $\Xi^0$ candidates are refit with the assumption that they originate at the decay vertex of their parent hyperon.

We identify primary hyperons by requiring that their decay vertex be displaced from the interaction point by $>2$~mm, and that their mass be within $5\sigma$ of its nominal value for $\Lambda^0$, and within $3\sigma$ of its nominal value for all other hyperons.  For those hyperons which decay into a $\Lambda^0$, each $\Lambda^0$ candidate is kinematically fitted to its nominal mass and is required to have a decay vertex at a greater distance from the interaction point than that of the primary hyperon.

The $\Lambda^0$ hyperons are reconstructed by kinematically fitting two oppositely charged tracks to a common vertex.  The higher momentum track is identified as a proton, and the lower momentum track is assumed to be a pion.  
The $\Sigma^+$ hyperons are reconstructed by combining protons with $\pi^0$ candidates.  The $\pi^0$ candidates are refit assuming that they come from the $\Sigma^+$ decay vertex and are combined with the proton to form the $\Sigma^+$ candidate. 

The $\Sigma^0$ hyperons are reconstructed by combining a $\Lambda^0$ candidate with a photon candidate. The photon candidate is required to have an energy greater than 50~MeV.  We select $\Sigma^0$ candidates only by requiring their masses to be within $3\sigma$ of the nominal $\Sigma^0$ mass~\cite{pdg}.

The $\Xi^-$ and $\Omega^-$ hyperons are reconstructed by combining a $\Lambda^0$ candidate with a charged track identified as $\pi^-$ and $K^-$, respectively. 

The $\Xi^0$ hyperons are reconstructed similarly to the $\Sigma^+$ hyperon, with the proton replaced by a $\Lambda^0$ candidate.   

Having identified single baryons, we construct the $e^+e^- \to B\overline{B}$ baryon--antibaryon pair events which are produced at rest.  To reconstruct these events, we select baryon-antibaryon pairs with a total momentum of $<50$~MeV.  If an event has multiple baryon--antibaryon pair candidates that pass these criteria, we take the pair with the smallest total momentum.   This eliminates backgrounds from events with additional particles, and yields an essentially background-free sample of events.

To determine the reconstruction efficiency of the above event selections, we generate Monte Carlo events using a GEANT-based detector simulation.  For the decay of $\psi(2S)$ to spin--1/2 baryon pairs ($\Lambda,\Sigma,\Xi$), we generate events with the expected angular distribution of $1+\cos^2 \theta$. 
For the spin--3/2 $\Omega^-$ hyperon, we generate events with the angular distribution  $[\sin\frac{\theta}{2}(1+3\cos\theta) + \cos\frac{\theta}{2}(1-3\cos\theta)]^2$ expected for spin~$1\to3/2+3/2$.

\begin{figure}[!tb]
\begin{center}
\includegraphics[width=3.5in]{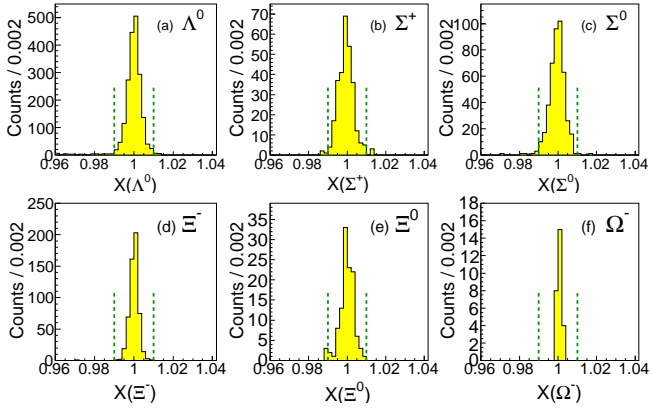}
\end{center}

\caption{Distributions of baryon--antibaryon events as function of, $X(B)\equiv(E(B)+E(\overline{B}))/\sqrt{s}$, in $\psi(2S)$ data.  The vertical lines indicate the signal region $X=0.99-1.01$.}
\label{fig:psi2sxb}
\end{figure}

\begin{figure}[!tb]
\begin{center}
\includegraphics[width=2.5in]{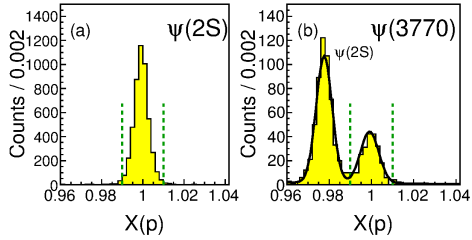}
\end{center}

\caption{Shows event distributions $X(p)\equiv [E(p)+E(\bar{p})]/\sqrt{s}$ for (a) $\psi(2S)\to p\bar{p}$, and (b) $p\bar{p}$ decays at $\psi(3770)$.  Allowed total momentum has been increased from $<50$~MeV to $<150$~MeV in order to show clearly the contribution from $\psi(2S)$ ISR excitation at $\sqrt{s}=3770$~MeV.}
\label{fig:proton}
\end{figure}

\begin{figure}[!tb]
\begin{center}
\includegraphics[width=3.3in]{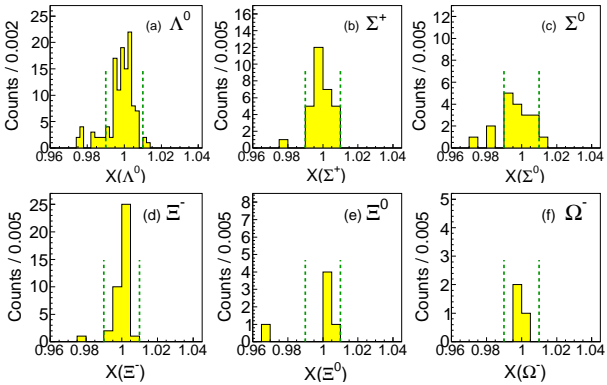}
\end{center}

\caption{Distributions of baryon-antibaryon scaled energy, $X(B)\equiv[E(B)+E(\overline{B})]/\sqrt{s}$, in $\sqrt{s}=3770$~MeV data.  The vertical lines indicate the signal region $X=0.99-1.01$.}
\label{fig:psi3770xb}
\end{figure}

As mentioned earlier, because the resonance decays $\psi(2S)\to B\overline{B}$ have large yields, they are best suited to illustrate the intermediate steps in our analysis.  The first step is to identify single hyperons as described before.  
The second step consists of constructing  baryon--antibaryon pairs. The distributions of the resulting $B\overline{B}$ pairs is shown in Fig.~\ref{fig:psi2sxb} for $\psi(2S)$ decays as a function of $X(B)\equiv[E(B)+E(\overline{B})]/\sqrt{s}$, which should peak at $X(B)=1$. Clear peaks are seen for all decays with essentially no background.  
We have studied large samples of generic MC data to determine potential backgrounds from other decays and find them to be $<0.1\%$ in the signal region, and therefore negligible.
We define the signal region as $X(B)=0.99-1.01$, with numbers of events in it as $N_\text{data}$.  We estimate the number of events, $N_\text{ff}$, due to form factor contribution under the peaks by extrapolating the form factor we measure at $\psi(3770)$, taking account of luminosity and efficiency differences, and the expected $(\sqrt{s})^{10}$ variation of the form factor.  We calculate the radiative correction, $(1+\delta)$, using the method of Bonneau and Martin~\cite{bonneaumartin}. We obtain $(1+\delta)=0.77$ within 1\% for all baryons at both the $\psi(2S)$ and $\psi(3770)$.  The Born cross sections are calculated as $\sigma_B=(N_\text{data}-N_\text{ff})/\epsilon_B\,\mathcal{L}(\psi(2S))\,(1+\delta)$, and the branching fractions as $\mathcal{B}(\psi(2S)\to B\overline{B}) = (N_\text{data}-N_\text{ff})/\epsilon_B N(\psi(2S))$.  The results are summarized in Table~\ref{tbl:psi2sbr}, including those for $\psi(2S)\to p\bar{p}$.  The first uncertainties in $\sigma_B$ and $\mathcal{B}$ are statistical, and the second uncertainties are estimates of systematic uncertainties as described below. 
Our results for the $\psi(2S)$ branching fractions are in agreement with the PDG averages~\cite{pdg} and previous small luminosity CLEO results~\cite{cleo-psi2sbb}, and have generally smaller errors.   
Fig.~\ref{fig:proton} illustrates the distributions of $p\bar{p}$ events for (a) $\psi(2S)\to p\bar{p}$, and (b) at the $\psi(3770)$.  In Fig.~\ref{fig:proton}(b), the ISR yield of $\psi(2S)\to p\bar{p}$ is also shown.

We apply the same event selections to the decays at the $\psi(3770)$ as we do for $\psi(2S)$ decays.  The $X(B)$ distributions for decays at the $\psi(3770)$ are shown in Fig.~\ref{fig:psi3770xb}.  Clear peaks are seen for each decay mode with yields ranging from 105 for $\Lambda^0\overline{\Lambda}^0$ to 3 for $\Omega^-\overline{\Omega}^+$.  The few events seen in the neighborhood of $X\approx0.98$ are consistent in number with being from the decay of the $\psi(2S)$ populated by initial state radiation (ISR). The number of events, $N_\text{ff}$, in the region $X(B)=0.99-1.01$, are used to calculate the cross sections as, $\sigma_0(e^+e^- \to B\overline{B}) = N_\text{ff}/(1+\delta)\epsilon_B\,\mathcal{L}(3770)$, where $\epsilon_B$ are the MC-determined efficiencies at $\sqrt{s}=3770$~MeV, $(1+\delta)=0.77$ is the radiative correction, and $\mathcal{L}(3770)=805$~pb$^{-1}$ is the luminosity at $\sqrt{s}=3770$~MeV.

\begin{figure}[!tb]
\begin{center}
\includegraphics[width=3.2in]{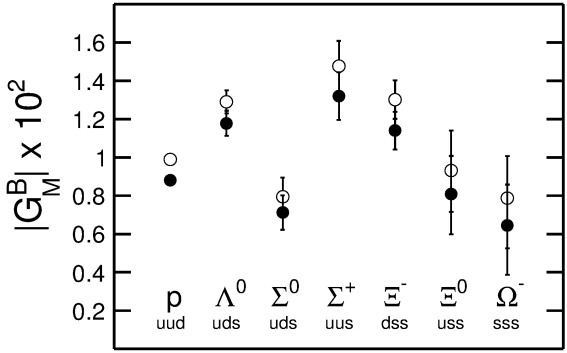}
\end{center}

\caption{Magnetic form factors $|G_M^B|\times10^2$ for proton and hyperons for $|Q^2|=14.2$~GeV$^2$.  The closed circles correspond to the assumption $|G_M^B|=|G_E^B|$, and the open circles to the assumption $|G_E^B|=0$.}
\label{fig:ffs}
\end{figure}

For the spin--1/2 baryons, the proton and the hyperons $\Lambda$, $\Sigma$, and $\Xi$, the well known relation between the cross sections and the magnetic form factor $|G_M^B(s)|$, and the electric form factor $|G_E^B(s)|$ is
\begin{equation}
\sigma_0^B = \left( \frac{4\pi\alpha^2\beta_B}{3s}\right) \left[ |G_M^B(s)|^2 + (2m_B^2/s)|G_E^B(s)|^2 \right]
\end{equation}
where $\alpha$ is the fine structure constant, $\beta_B$ is the velocity of the baryon in the center-of-mass system, and $m_B$ is its mass.  The statistics of the present measurements do not allow us to determine $|G_M^B|$ and $|G_E^B|$ separately.  We therefore evaluate $|G_M^B(s)|$ under two commonly used extreme assumptions,  $|G_E^B(s)|/|G_M^B(s)|=0$, and 1.  The results corresponding to $|G_E^B|=|G_M^B|$ are shown in Table~\ref{tbl:hypff}.  
The efficiencies for the $|G_M|$ and $|G_E|$ components are determined assuming $1+\cos^2\theta$ and $\sin^2\theta$ angular distributions, respectively.
In Fig.~\ref{fig:ffs}, we also plot the values of $|G_M^B|$ derived with the assumption $|G_E^B|=0$. They are between 10\% and 15\% larger than those obtained with the assumption $|G_E^B|=|G_M^B|$.

For the spin--3/2 $\Omega^-$, there are four form factors, $G_{E0}$, $G_{E2}$, $G_{M1}$, and $G_{M3}$~\cite{omegaff}.  Following K\"orner and Kuroda~\cite{kornerkuroda}, Eq.~2 is valid if it is understood that $|G_M^B|$ includes the contributions of both magnetic quadrupole and octopole form factors, and $|G_E^B|$ includes the contributions of both electric dipole and quadrupole form factors.

We evaluate systematic uncertainties due to various sources for each hyperon pair, and add the contributions from the different sources together in quadrature.  The uncertainties due to particle reconstruction are 1\% per charged particle, 2\% per $\gamma$, 2\% per $\pi^0$, and 1\% per hyperon.  There are additional uncertainties of 2\% per $p$ and $K$ due to the use of RICH and $dE/dx$ information.  Other systematic uncertainties are 2\% in $N(\psi(2S))$, 1\% in $\mathcal{L}(\sqrt{s}=3770)$, and 0.2\% in the radiative correction.  These systematic uncertainties total 6.1\% for $\Lambda^0$, 7.3\% for $\Sigma^0$, 6,4\% for $\Sigma^+$, 7.5\% for $\Xi^-$, 7.3\% for $\Xi^0$, and 10.2\% for $\Omega^-$.

\begin{table}[!tb]
\setlength{\tabcolsep}{4pt}
\begin{tabular}{l|ccccc}
\hline \hline
$\mathcal{B}$ & $N_\text{data}$ & $N_\text{ff}$ & $\epsilon_B$ (\%) & $\sigma_B$ (pb) & $\mathcal{B}\times10^4$  \\
\hline
$p$         & $4475(78)$ & 16.0(10) & 63.1  & $196(3)(12)$ & $3.08(5)(18)$  \\
$\Lambda^0$ & $1901(44)$ &  7.9(7)  & 20.7  & $247(6)(15)$  & $3.75(9)(23)$  \\
$\Sigma^0$  & $ 439(21)$ &  1.1(3)  & 7.96  & $148(7)(11)$  & $2.25(11)(16)$ \\
$\Sigma^+$  & $ 281(17)$ &  2.2(3)  & 4.54  & $165(10)(11)$ & $2.51(15)(16)$ \\
$\Xi^-$     & $ 548(23)$ &  2.9(4)  & 8.37  & $176(8)(13)$  & $2.66(12)(20)$ \\
$\Xi^0$     & $ 112(11)$ &  0.4(2)  & 2.26  & $135(13)(10)$ & $2.02(19)(15)$ \\
$\Omega^-$  & $  27(5)$  &  0.2(1)  & 2.32  & $31(6)(3)$    & $0.47(9)(5)$   \\

\hline \hline
\end{tabular}

\caption{Cross section and branching fraction results for $\psi(2S)\to B\overline{B}$.}  
\label{tbl:psi2sbr}
\end{table}

\begin{table}[!tb]
\begin{center}
\begin{tabular}{lc|cccc}
\hline \hline

$B$ & $\mu_B$ & \multicolumn{1}{c}{$N_\text{ff}$} & $\epsilon_B$, \% & $\sigma_0^B$, pb & $|G_M^B|\!\times\!10^2$   \\
\hline
$p$         &  2.79 & $215(15)$ & $71.3$ & $0.46(3)(3)$ & $0.88(3)(2)$ \\
$\Lambda^0$ & $-0.61$ & $105(10)$  & $21.1$ & $0.80(8)(5)$  & $1.18(6)(4)$  \\
$\Sigma^0$  & (0.79)  & $ 15(4)$  & $8.36$ & $0.29(7)(2)$  & $0.71(9)(3)$  \\
$\Sigma^+$  &  2.46   & $ 29(5) $  & $4.68$ & $0.99(18)(6)$ & $1.32(13)(4)$ \\
$\Xi^-$     & $-0.65$ & $ 38(6) $  & $8.69$ & $0.71(11)(5)$ & $1.14(9)(4)$  \\
$\Xi^0$     & $-1.25$ & $ 5^{+2.8}_{-2.3}$ & $2.30$ & $0.35^{+0.20}_{-0.16}(3)$ & $0.81(21)(3)$ \\
$\Omega^-$  & $-2.02$ & $ 3^{+2.3}_{-1.9}$ & $2.94$ & $0.16^{+0.13}_{-0.10}(2)$ & $0.64^{+0.21}_{-0.25}(3)$ \\

\hline \hline
\end{tabular}
\end{center}

\caption{Results for proton and hyperon form factors at $|Q^2|=14.2$~GeV$^2$, assuming $|G_E^B|=|G_M^B|$. The known uncertainties in $\mu_B$ are all less than $\pm 2\%$. The magnetic moment for $\Sigma^0$ is based on the PDG fit to quark model predictions for the hyperons~\cite{pdg}.}  

\label{tbl:hypff}
\end{table}

Since no modern theoretical predictions for timelike form factors of hyperons at large momentum transfers exist, we can only discuss our experimental results qualitatively.  Following are the main observations:
\begin{itemize}
\item[(a)] The $e^+e^-\to\gamma^*\to B\overline{B}$ cross sections in Table~\ref{tbl:hypff} are 150 to 500 times smaller than the resonance cross sections in Table~\ref{tbl:psi2sbr}, as was expected on the basis of Eq.~1.   Clearly, larger statistics measurements of the form factors would be highly desirable.

\item[(b)] As illustrated in Fig.~\ref{fig:ffs},  except for $|G_M(\Sigma^0)|$, the measured values of $|G_M^B|$ vary by approximately a factor two. The pattern of $SU(3)$ breaking is not obvious, except that we do observe that there is monotonic decrease in the form factors as the number of strange quarks increases from one in the $\Sigma^+$, to two in the $\Xi$, to three in the $\Omega^-$.

\item[(c)] It is common practice to quote spacelike form factors for protons as $|G^p_M(s)/\mu_p|$, based on normalization at $|Q^2|=0$.  For timelike momentum transfers, no such relation between $\mu_B$ and $|G_M^B|$  is expected, and none appears to exist, with $\mu_B$ as listed in Table~\ref{tbl:hypff}.

\end{itemize}
The most significant result of the present measurements is that $|G_M(\Lambda^0)|$ is a factor $1.66(24)$ larger than $|G_M(\Sigma^0)|$, although the $\Lambda^0$ and $\Sigma^0$ have the same $uds$ quark content. We note that the $\Sigma^0$ and $\Lambda^0$ differ in their isospin, with $I(\Sigma^0)=1$, and $I(\Lambda^0)=0$.  Since only up and down quarks carry isospin, this implies that the pair of up/down quarks in the $\Lambda^0$ and $\Sigma^0$ have different isospin configurations.  This forces different spin configurations for the $ud$ quarks in the $\Lambda^0$ and $\Sigma^0$.  In the $\Lambda^0$ the $ud$ quarks have antiparallel spins coupled to $S=0$, whereas in the $\Sigma^0$ they couple to $S=1$.  The spatial overlap in the $S=0$ configuration in the $\Lambda^0$ is stronger than in the $S=1$ configuration in the $\Sigma^0$. 
This interpretation is further supported by the fact that in contrast to $G_M(\Sigma^0)$, $G_M(\Sigma^+)=1.32(13)$ is essentially equal to $G_M(\Lambda^0)=1.18(7)$.  Unlike the $S=1$ coupled $ud$ quarks in $\Sigma^0$, in $\Sigma^+$ the overall space, spin, and isospin antisymmetrization forces to the two like $uu$ quarks to $S=0$, like the $ud$ quarks in isospin zero $\Lambda^0$ leading to $G_M(\Sigma^+)\approx G_M(\Lambda^0)$. Our measurements at large $|Q^2|$ are particularly sensitive to such short range correlations.  

It is interesting to note that in a measurement of production of $\Lambda^0$ and $\Sigma^0$ with polarized photons, Bradford~et~al.~\cite{bradford} had observed large differences in polarization observables of $\Lambda^0$ and $\Sigma^0$, and without explicitly attributing them to diquark correlations, had noted that ``the differences were perhaps not surprising since the spin structure of the  $\Sigma^0$ and $\Lambda$ are different.'' 

Recently, Jaffe~\cite{jaffe} and Wilczek~\cite{wilczek} have emphasized the importance of diquark correlations in low-energy QCD dynamics, and have pointed out that for the non-strange quarks the favorable diquark configuration with attraction is the spin-isospin singlet, making what Wilczek calls a ``good'' diquark in the $\Lambda^0$ as opposed to the repulsive spin-isospin triplet configuration in the $\Sigma^0$.  This results in a significantly larger cross section for the formation of the $\Lambda^0$ than $\Sigma^0$, as anticipated by Selem and Wilczek~\cite{wilczek}.  We measure $\sigma(\Lambda^0)/\sigma(\Sigma^0)\approx3$, and this results in the factor 1.66 larger form factor for the $\Lambda^0$ than $\Sigma^0$. We find that our observation of the large difference between the form factors of the $\Lambda^0$ and $\Sigma^0$ can be attributed to the ``good'' diquark correlation in the $\Lambda^0$.

This investigation was done using CLEO data, and as members of the former CLEO Collaboration we thank it for this privilege.  This research was supported by the U.S. Department of Energy.  The authors also wish to thank Professors G.~Miller, S.~Brodsky, and W.~Roberts for helpful comments.

\bibliographystyle{model1a-num-names}

\begin{thebibliography}{99}
\bibitem{fn1} We note for completeness that we do not measure the form factor for $\Sigma^-$ because its major decay mode ($\Sigma^- \to n \pi^-$) is extremely difficult to reconstruct.
\end{thebibliography}

\end{document}